\documentclass{article}

\usepackage{amsmath}
\usepackage{arxiv}

\usepackage[utf8]{inputenc} 
\usepackage[T1]{fontenc}    
\usepackage{hyperref}       
\usepackage{url}            
\usepackage{booktabs}       
\usepackage{amsfonts}       
\usepackage{nicefrac}       
\usepackage{microtype}      
\usepackage{lipsum}
\usepackage{graphicx}
\graphicspath{ {./images/} }

\title{Feature Identification in Complex Fluid Flows by Convolutional Neural Networks}

\author{
 Shizheng Wen \\
  Department of Mathematics\\
  ETH Zurich\\
  8092 Zürich, Switzerland \\
  \texttt{shiwen@student.ethz.ch} \\
   \And
 Michael W. Lee \\
  NASA Langley Research Center\\
  Hampton, VA 23666, USA\\
  \texttt{michael.w.lee@pm.me} \\
  \And
 Kai M. Kruger Bastos \\
  Rivian\\
  San Francisco, CA 94080, USA \\
  \texttt{kai.mbali.kruger.bastos@duke.edu} \\
  \And
 Earl H. Dowell \\
  Department of Mechanical Engineering and Materials Science\\
  Duke University\\
  Durham, NC 27708, USA \\
  \texttt{earl.dowell@duke.edu} \\
}

\begin{document}
\maketitle
\begin{abstract}
Recent efforts have shown machine learning to be useful for the prediction of nonlinear fluid dynamics. Predictive accuracy is often a central motivation for employing neural networks, but the pattern recognition central to the network’s function is equally valuable for purposes of enhancing our dynamical insight into confounding dynamics. In this paper, convolutional neural networks (CNNs) were trained to recognize several qualitatively different subsonic buffet flows over a high-incidence airfoil, and a near-perfect accuracy was performed with only a small training dataset. The convolutional kernels and corresponding feature maps, developed by the model with no temporal information provided, identified large-scale coherent structures in agreement with those known to be associated with buffet flows. An approach named Gradient-weighted Class Activation Mapping (Grad-CAM) was then applied to the trained model to indicate the importance of these feature maps in classification. Sensitivity to hyperparameters including network architecture and convolutional kernel size was also explored, and results show that smaller kernels are better at coherent structure identification than are larger kernels. A long-short term memory CNN was subsequently used to demonstrate that with the inclusion of temporal information, the coherent structures remained qualitatively comparable to those of the conventional CNN. The coherent structures identified by these models enhance our dynamical understanding of subsonic buffet over high-incidence airfoils over a wide range of Reynolds numbers.
\end{abstract}


\section{Introduction}
\label{sec: 1}
In recent years, with the development of high-performance computing architectures and experimental measurement capabilities~\cite{zhuang2019fractal,mustafa2019amplification}, fluid researchers are able to obtain high precision and high-resolution spatiotemporal data of large-scale fluid simulations and experiments. Too, the advancement of sophisticated algorithms and the abundance of open source software enables researchers to apply machine learning (ML) to address many challenges~\cite{li2021recent,cai2022physics,kochkov2021machine,zhang2021artificial,brunton2020machine,li2021recent}. Turbulence modeling and, more generally, nonlinear fluid dynamics has been one proving ground for neural networks~\cite{duraisamy2019turbulence,kou2021data}.

For nonlinear fluid flow regimes, incorporating domain knowledge into learning algorithms has been demonstrated to be feasible. Data-driven turbulence modeling presents promising extensions to more conventional numerical system closure techniques and are therefore of significant value for engineering applications~\cite{duraisamy2019turbulence,panda2022evaluation,zhu2022one}. For example, Tracey et al.~\cite{tracey2015machine} successfully reproduced the Spalart-Allmaras model by replacing the deliberately removed source term with machine-learned functional forms. Duraisamy’s group also pursued efforts in data-driven turbulence modeling~\cite{zhang2015machine,singh2017machine,singh2016using,parish2016paradigm,duraisamy2015new,duraisamy2021perspectives} with encouraging results; they were able to infer functional forms of modeling discrepancies by using inverse modeling, and then were able to reconstruct the patterns with ML for incorporation into turbulence model source terms. Xiao’s group emphasized the physical constraint of Reynolds stress and proposed the concept of a physics-informed machine learning approach~\cite{wang2017physics,wu2017priori}. Ling et al.~\cite{ling2016reynolds} reconstructed the mapping relations between field variables and the Reynolds stress anisotropy tensor, and replaced the turbulence model with a tensor-based neural network. Zhu et al.~\cite{zhu2019machine} completely replaced the Reynolds stress transport equations with neural networks and then constructed a mapping function between the turbulent eddy viscosity and the mean flow variables.

In addition to data-driven turbulence modeling, characteristics of deep learning (DL) algorithms\cite{schmidhuber2015deep,donahue2015long,shi2015convolutional} like the convolutional neural network (CNN) and the long-short term memory network (LSTM) provide tools for researchers to evaluate the temporal and spatial patterns in data. For example, Ye et al.~\cite{ye2020flow} applied CNN to predict the pressure coefficient on a cylinder from velocity distributions in its wake flow. Zhang’s group explored many potential applications of deep learning modeling for unsteady aerodynamics and aeroelasticity~\cite{kou2021data}. Zhang et al.~\cite{zhang2018application} trained multiple linked CNNs to learn the lift coefficients of an airfoil with a variety of shapes in multiple flow regimes. Guo et al.~\cite{guo2016convolutional} proposed a convolutional encoder-decoder approach that can predict steady velocity and pressure fields. Bhatnagar et al.~\cite{bhatnagar2019prediction} improved the computational efficiency of this effort by sharing encoder-decoder layers. Mohan et al.~\cite{mohan2018deep} built a deep learning approach to reduced order modeling (ROM) for isotropic turbulent flows by replacing Galerkin projection with LSTM neural networks.

The success of the aforementioned works indicates the encouraging prospects of ML in the fluid mechanics. Furthermore, it also adequately demonstrates that ML can extract intrinsic flow features for use in establishing a nonlinear mapping relationship with the desired output~\cite{ye2020flow,colvert2018classifying}. However, previous research efforts have focused primarily on the accuracy of predictive variables, without studying closely the information hidden inside the learning model itself. Additionally, the more dynamically motivated efforts have utilized flows with high levels of symmetry, e.g. isotropic turbulence. In this paper, we employ conventional ML implementations to identify coherent structures in a flow important to engineering applications: one over an airfoil at a high angle of attack, where subsonic buffet is known to occur. The coherent structures associated with buffet, identified through feature maps associated with the convolutional kernels, align with and expand upon the previous mathematical and physical insights for the problem despite being identified entirely by the ML algorithms. A side-effect of this coherent structure identification was a near-perfect flow identification capability with only small training dataset, where the neural network learned quickly how to recognize qualitatively different flow regimes from individual temporal snapshots.

The remainder of this paper is organized as follows. In section~\ref{sec: 2}, the problem of subsonic buffet and the computation of the flow data provided to the neural networks are discussed. In section~\ref{sec: 3}, results from the CNN architecture are presented and sensitivity to certain hyperparameters are discussed. In section~\ref{sec: 4}, results from the LSTM architecture are presented with the conclusion that little difference is observed between the CNN and LSTM coherent structures.

\section{Problem Formulation}
\label{sec: 2}
Flow over a NACA 0012 airfoil was computed via direct numerical simulation (DNS) at a constant incidence angle of 40 degrees and temporally constant Reynolds numbers ranging from 100 to 1,000,000. The Mach number remained constant in all flows at approximately 0.05. These conditions elicited a spectrum of temporally fluctuating flows over the airfoil: some periodic, some quasi-periodic, and some chaotic. These unsteady, lift-generating flows over the airfoil are collectively known as subsonic buffet.
\subsection{Subsonic Buffet}
\label{sec: 2.1}
The buffet phenomenon in subsonic open flows has received recent attention~\cite{tang2014experimental,besem2016vortex}. Vortex shedding similar to bluff body flows is observed in high-incidence flows around stationary airfoils~\cite{tang2014experimental,besem2016vortex,zhou2018buffeting}. Lift coefficients are one way to characterize these time-periodic dynamic instabilities. A scaling analysis of the Navier-Stokes equations~\cite{jaworski2012scaling} demonstrates that both the peak-to-peak oscillating lift and reduced frequency in buffeting flows are of order unity. Consequently, while the kinematics of the flow are not yet fully understood, the buffet dynamics are of large-enough scale to dramatically affect the airfoil performance. It is these large-scale structures that are identified through ML in this research.
\subsection{DNS Flow Simulations}
\label{sec: 2.2}
The neural networks central to this research were presented with snapshots of fluid flows around an airfoil across a range of Reynolds numbers. The subsonic flows exhibited characteristics of buffet, in which the air foil’s lift coefficient oscillated in time but had a positive mean value. The method of simulating such a flow is an active area of research; readers are directed to Ref.~\cite{bastos2020computational} for further reading. Unsteady reynolds-averaged navier-stokes (URANS) simulations have been shown to present reduced frequencies comparable to those of experiments, but the amplitudes of the oscillations were more sensitive to the choice of closure model. The URANS simulations also failed to exhibit buffet at all at some lower Reynolds numbers which were known from experiments to yield buffet.
\begin{figure}
\centering
\includegraphics[width=0.5\textwidth]{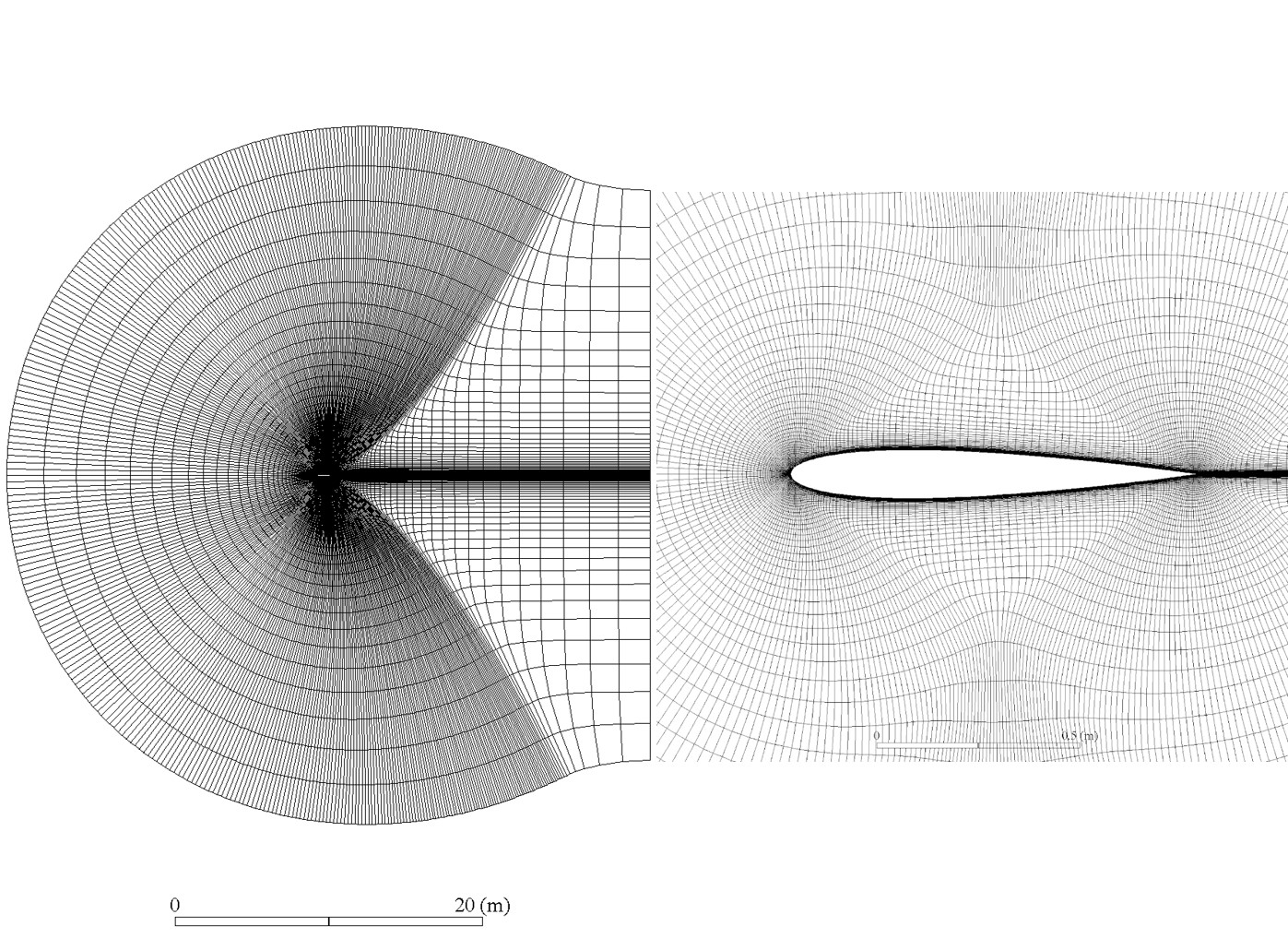}
\caption{\label{fig:1} Full view (left) and close-up view (right) of the 2D airfoil grid used for DNS simulations.}
\end{figure}

Recent efforts utilizing “unresolved direct numerical simulation” (UDNS), also known as “full Navier-Stokes” (Full NS), have provided a numerically tractable method of simulating such flows with large-scale behavior comparable to that of experiments. In particular, the reduced frequencies of the lift and drag coefficients were computed to be within 7\% of the experimental results at the same high Reynolds numbers~\cite{bastos2020computational}.

This method is known to be only conditionally stable in time. Too, it is understood that this method does not guarantee, nor does it seek, accuracy in subgrid-scale dynamics. However, for unsteady aerodynamic phenomena dominated by large-scale flow structures, e.g. subsonic buffet, UDNS presents a compromise between computational cost and agreement with experiment. It was thus utilized in this study to generate a snapshot database for use by the neural networks, which themselves only studied large-scale flow structures. The pattern recognition of the neural network was not contingent upon fully resolved small-scale flow features, and thus the purpose of this study did not require resolution down to the Kolmogorov scale.
\begin{figure}
\centering
\includegraphics[width=0.8\textwidth]{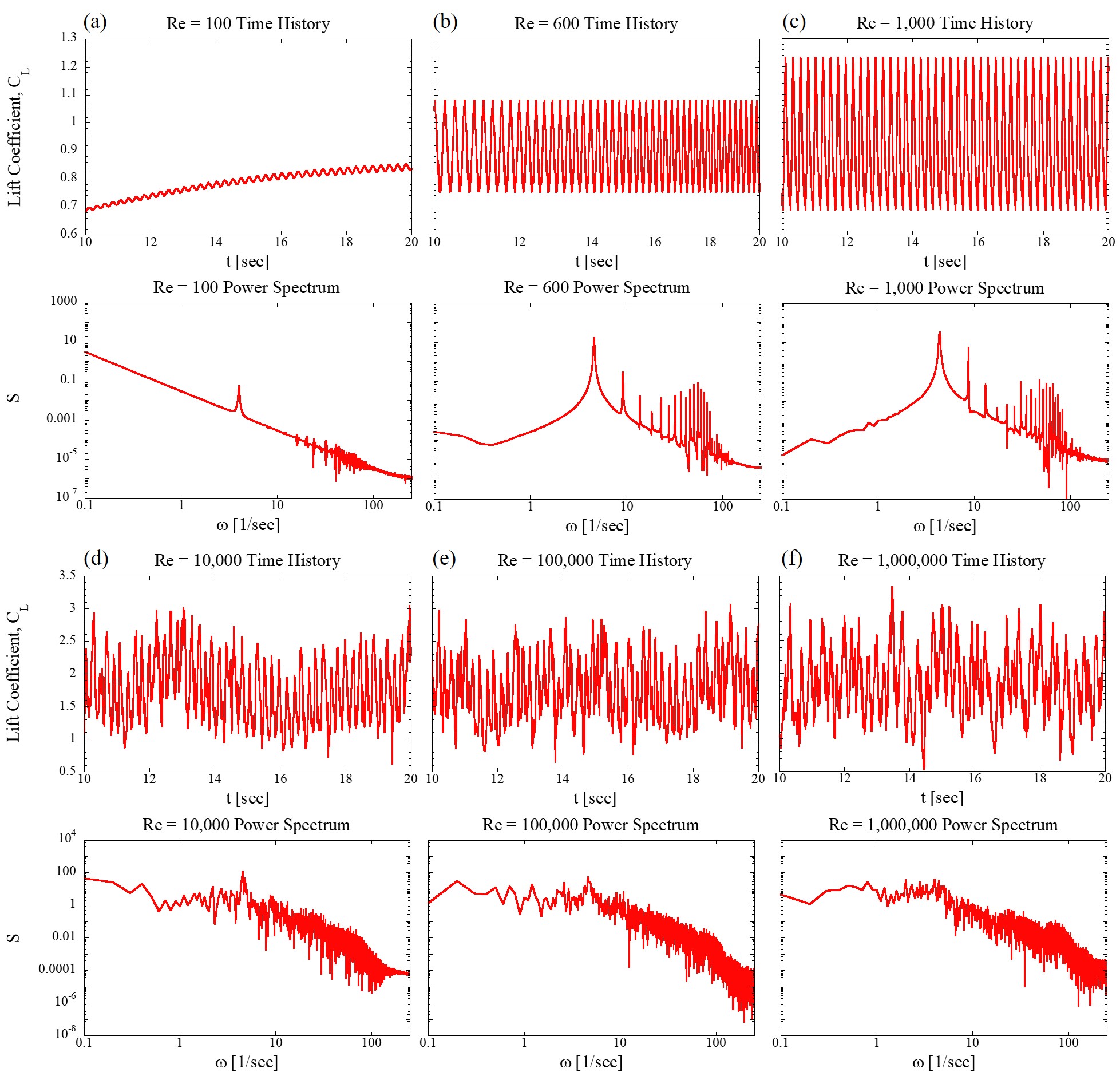}
\caption{\label{fig:2} Response of lift coefficient changing with time and corresponding power spectrum at Reynolds numbers of 100, 600, 1,000, 10000, 100,000 and 1,000,000.}
\end{figure}

Flows were thus simulated in time about a symmetric NACA 0012 airfoil at several Reynolds numbers: 100, 600, 1,000, 10,000, 100,000 and 1,000,000. All six simulations started with a control volume at rest and converged to a statistically steady state; a constant time step of 0.002 seconds was employed. The incidence angle was held at 40 degrees and Mach number was held constant at approximately 0.05: strongly within the incompressible limit. Simulations were performed with a truncated NASA grid~\cite{rumsey2010description}, with 257 airfoil surface points, as illustrated in Fig.~\ref{fig:1}. The grid extends 20 chord lengths in all directions, which was determined to be adequate to resolve far-field behavior~\cite{kuhnen2018destabilizing}. The Navier-Stokes equations were solved directly in the ANSYS Fluent software package to generate all flow data used in this study. At a Reynolds number of 87,000, at which the flow parameters matched those of Tang and Dowell’s experimental study~\cite{tang2014experimental}, the lift and drag coefficient reduced-frequencies were observed to be in good agreement between the DNS simulation and experimental results.

Figure~\ref{fig:2} shows steady-state time histories and power spectra of the lift coefficient (CL) as computed at different Reynolds numbers. By employing the heuristic flow characterization developed by Wiebe and Virgin~\cite{wiebe2012heuristic} and in agreement with existing literature~\cite{rossi2018multiple}, each Reynolds number is associated with qualitatively different flow regimes: periodic flow, quasi-periodic flow and chaotic flow, as shown in Table~\ref{tbl1}. All simulations were performed with a time step of 0.002 seconds. Consequently, the 2,000 snapshots at Re = 100, for example, correspond to the flow as simulated between 16 and 20 dimensional seconds after the simulation was started from rest. Recent work by one of the authors~\cite{bastos2020computational} determined that there was weak dependence on grid resolution for large ranges of Reynolds number when seeking the lift and drag coefficient reduced frequencies. Consequently, the reduced frequencies are assumed to be dominated by large-scale features. It is these local but large-scale flow features that this study has identified by using ML.
\begin{table}
 \caption{High-fidelity DNS snapshots provided to the neural networks, classified by Reynolds number as qualitatively different flow regimes.}
  \centering
  \begin{tabular}{llll}
  \toprule
   Reynolds Number & Sampling region & Flow regime \\
    \midrule
100 & 8000-10,000 & Periodic flow \\
600 & 8000-10,000 & Quasi-periodic flow \\
1,000 & 8000-10,000 & Quasi-periodic flow \\
10,000 & 5000-10,000 & Chaotic flow \\
100,000 & 5000-10,000 & Chaotic flow \\
1,000,000 & 5000-10,000 & Chaotic flow \\
    \bottomrule
  \end{tabular}
  \label{tbl1}
\end{table}

\subsection{Data Preprocessing}
\label{sec: 2.3}
\begin{figure}
\centering
\includegraphics[width=0.7\textwidth]{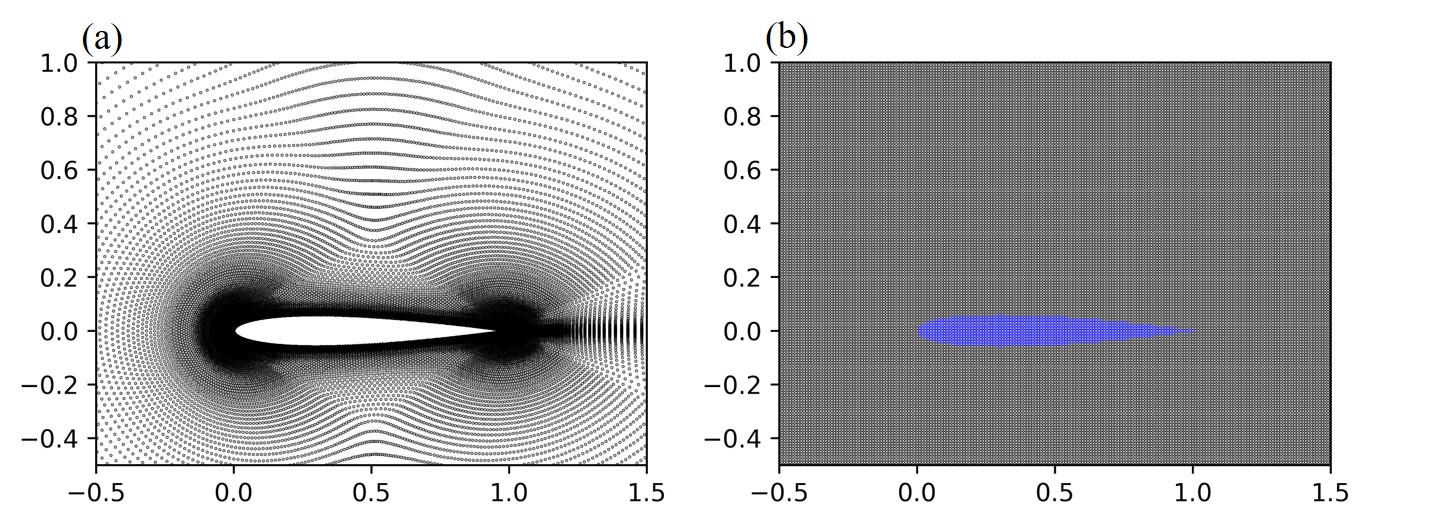}
\caption{\label{fig:3} DNS grid subsection (a) and the Cartesian grid (b) provided to the neural networks.}
\end{figure}

Because the lift coefficient is only determined by the airfoil surface pressures and the near-wall flow dynamics are indicative of the qualitative flow behavior, the flow region for ML consideration was confined to the area shown in Fig.~\ref{fig:3}(a). It is clear that the DNS grid structure is not Cartesian. For easier application of a square Cartesian convolutional kernel to the flow snapshots, the spatial grid was thus adjusted to also be Cartesian. The subsequent 200-by-150 pixel Cartesian grid is shown in Fig.~\ref{fig:3}(b). Flow values within the airfoil (the blue area in Fig.~\ref{fig:3}(b)) area were set to zero. Splines defined by the NACA airfoil standard~\cite{kurtulus2015unsteady} were used to determine what points were within the airfoil area.

\section{Convolutional Neural Network Implementation}
\label{sec: 3}
A convolutional neural network was employed to identify the large-scale coherent structures associated with the qualitatively different manifestations of buffet. All neural networks in this paper were constructed within the Google tensorflow environment. A summary of the mathematics of the CNN are presented in the supplementary material. With only small training dataset, the trained model was able to virtually unilaterally discern the qualitative flow state based on a single arbitrary temporal snapshot. Too, the model’s kernels – depending somewhat on the kernel size – were able to identify all dynamically significant large-scale flow features with no physical insight provided a priori. Gradient-weighted class activation mapping (Grad-CAM) was subsequently used to indicate the importance of these identified flow features in the presence of chaotic flow.
\subsection{Formulation}
\label{sec: 3.1}
The artificial neural networks (ANNs) first became prominent in the field of artificial intelligence in the 1980s~\cite{mcclelland1987parallel}. They are straightforward abstractions of biological neurons, realized as elements in an artificial network like a program or silicon-based circuits~\cite{demuth2014neural}. They present advantages in learning nonlinear and complex relationships, good generalization of unseen data and simplicity of input data formatting. For these and other reasons, ANNs continue to play an important role in modern ML methods. However, due to the fully-connected nature of neurons in two adjacent layers, traditional ANNs will generate a large number of parameters – many of which will likely be superfluous – when adding additional layers or increasing the number of neurons in each layer. This in turn often gives rise to overfitting of the model as the optimization procedure converges. With multidimensional data, even something as simple as static images, interpreting them as 1D vectors (as is necessary in traditional ANNs) will 1) cause an explosion of parameters and 2) destroy the spatial relationships which define the multidimensional inputs. To address these limitations, a specific ANN architecture was developed: the convolutional neural network (CNN)~\cite{gu2018recent}. LeCun et al.~\cite{lecun1998gradient} firstly proposed the layout of LeNet-5 in 1988 and was notably successful in a handwritten digit recognition task. Subsequently, many architectures derivative of the LeNet-5 further explored the image recognition problem, with much success~\cite{simonyan2014very,krizhevsky2012imagenet,szegedy2015going}.
\begin{figure}
\centering
\includegraphics[width=0.8\textwidth]{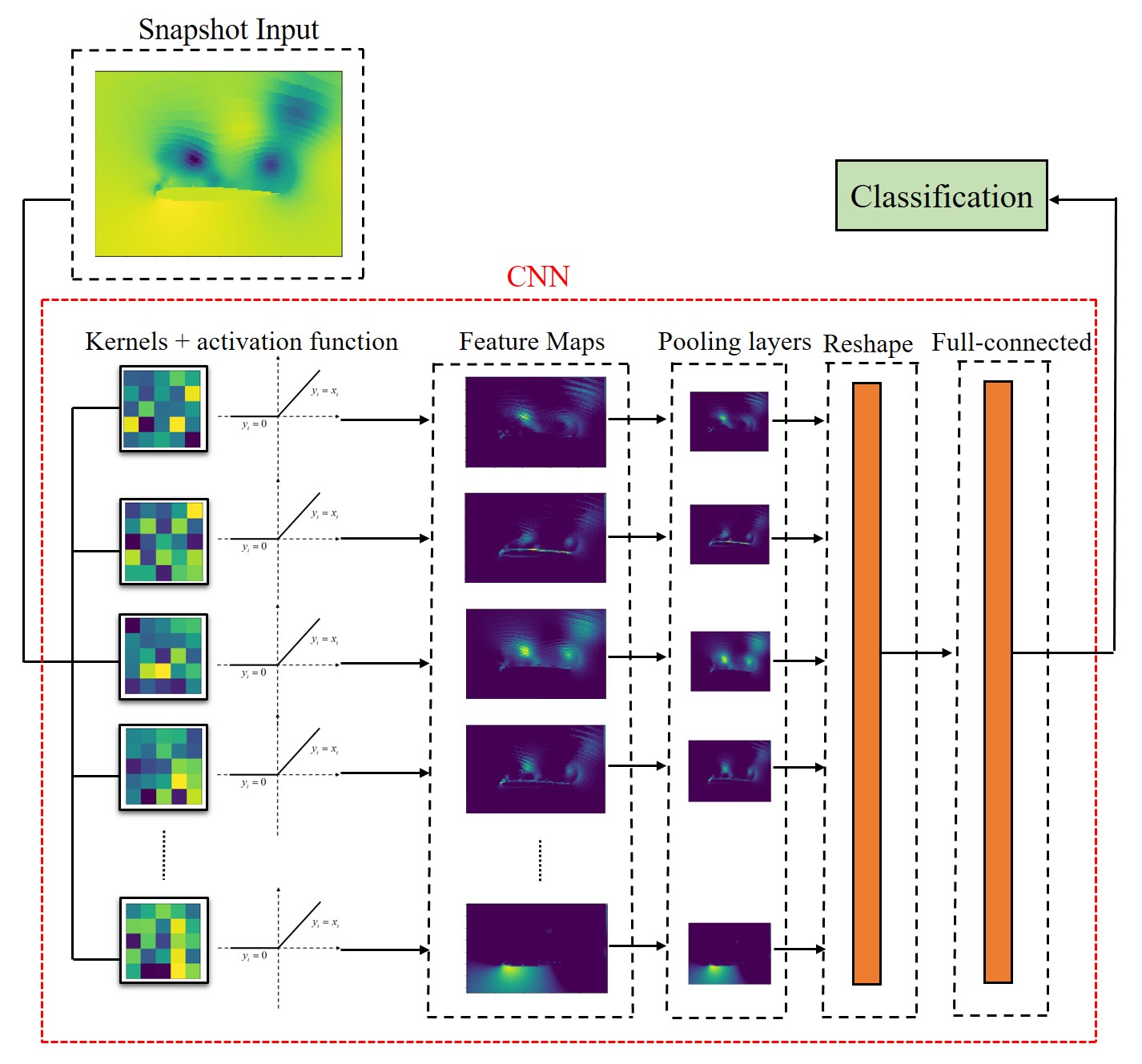}
\caption{\label{fig:4} Architecture of the conventional CNN employed for buffet flow classification.}
\end{figure}

CNNs derive their utility from seeking patterns at multiple scales through the process of convolving small kernels – perhaps five pixels square – with a much larger input signal – perhaps a grayscale image of 100 pixels square. CNNs thus are particularly successful when the input data can be decomposed into some form of hierarchical basis representation; and this is often labeled as automatic feature extraction~\cite{beck2019deep}. The optimization problem stems from finding kernels which identify the most fundamental patterns in the signal, which can then combine at higher levels to yield larger-scale patterns which in turn reconcile with provided data labels. For example, a face may be identified in an image at a high layer of the CNN because eyes and mouths are identified at a lower CNN layer, which are themselves identified at the lowest layer by kernels which design themselves to recognize edges and angles in the provided image. While the problem remains one of optimizing a loss function within some very high-dimensional parameter space, the convolution of both the kernels with the signals and the kernels at different layers significantly reduces the computational cost of training such a model when compared to the traditional ANN architecture. The former is known as local connectivity and the latter is known as weight sharing.

The main purpose of adding a convolutional layer to a ML model is thus to exploit the low-dimensional, high-level representation of the input data. A fully connected layer can then be employed to build the mapping relationship between these high-level representations and predictive variables. In this paper, we trained a CNN to achieve a simple classification task for the fluid problem discussed in section~\ref{sec: 2.2}. The kernels and corresponding feature maps were then extracted to study the flow features which the model identified. A summary of the CNN architecture used in this work is shown in Fig.~\ref{fig:4}. Methods of regularization, dropout, exponential learning rate decay and moving average were used in order to avoid overfitting and improve the robustness of the model. Hyperparameters of the CNN, which came about through sensitivity studies and review of similar ML models in the literature, are summarized in Table~\ref{tbl2}.
\begin{table}
\caption{CNN hyperparameters.}\label{tbl2}
\centering
\begin{tabular}{llll}
\toprule
Hyperparameter & Value & Hyperparameter & Value\\
\midrule
\textbf{Architecture of CNN} &  & \textbf{Optimization of CNN} & \\
Size of square convolutional kernel & 3/5/10/20 & Optimizer & Adam \\
Number of convolutional kernel & 10 & Batch size & 100 \\
Activation function & Relu & Training steps & 10,000 \\
Stride for convolution & 1 & Learning rate base& 0.0005 \\
Stride for pooling & 2 & Learning rate decay & 0.99 \\
Padding for convolution & Yes & Regularization & 0.0001 \\
Padding for pooling & Yes & Moving average decay & 0.99 \\
Number of units in fully connected layer & 200 & &  \\
Dropout ratio for fully connected layer & 0.5 &  &  \\
\bottomrule
\end{tabular}
\end{table}

Only information about the pressure field was provided to the neural network in this study. It was observed that the (normalized) pressure information was highly dominant in the training process even when the (normalized) velocity information was also provided to the neural network. This in itself is a significant observation, as subsonic buffet is known to stem, largely but possibly indirectly, from airfoil surface pressure gradients; the neural network likewise concluded from its observations that only the pressure fields distinguished qualitatively different types of buffet. It was clear that the neural networks were not equally valuing the velocity information for three reasons. Firstly, the flattened convolutional kernels and the corresponding feature maps did not change significantly in networks provided with the velocity and pressure information versus just the pressure information. Secondly, the kernels associated with the velocity information did not identify coherent structures in the same way that did the pressure kernels. Thirdly, a neural network provided with only velocity information did not yield dynamically significant coherent structures nearly as well as did the networks provided with only pressure information.

\subsection{Results and Discussion}
\label{sec: 3.2}
A selection of 600 out of the total 21,000 snapshots (detailed in Table~\ref{tbl3}) were used to train 4 CNNs, each of which had different kernel sizes as outlined in Table~\ref{tbl2}. The remaining dataset (20,400 snapshots) was employed to test the trained models. The training dataset was constructed to ensure each qualitatively different flow state (see Table~\ref{tbl1}) was represented by the same number of snapshots. The training data set was very small and accounted for the different buffet states but not all of the simulated Reynolds numbers. However, the trained models performed with high accuracy (over 0.95) for the test dataset. This was a remarkably small training dataset for the relatively high classification accuracy especially at the highest Reynolds number. Of note is that no training data was provided from the highest and most turbulent Reynolds number, whose flow was quantitatively very different from the other chaotic Reynolds numbers simulated, yet the trained model still recognized those snapshots as chaotic. This result indicates that the CNN adequately mapped the relationship between snapshots and qualitative flow states. What’s more, though the chaotic flow is temporally very complex, its coherent structures (as identified by the CNN) are not temporally sensitive. Such a conclusion is in agreement with the separability assumptions employed in empirical modeling methods like Galerkin POD-based reduced-order models~\cite{taira2017modal}.
\begin{table}
\caption{Selection of training dataset.}\label{tbl3}
\centering
\begin{tabular}{llll}
\toprule
Reynolds Number & Sampling region & Flow type\\
\midrule
100 & 8000-10,000 & Periodic flow \\
600 & 8000-10,000 & Quasi-periodic flow \\
10,000 & 5000-10,000 & Chaotic flow \\
\bottomrule
\end{tabular}
\end{table}

The few snapshots at the highest Reynolds number which were not identified correctly were observed to be qualitatively very similar to lower Reynolds number flows; Consequently, the coherent structures were correctly identified but the occasional appearance of simpler structures in the chaotic flow led to minor deviations from perfect classification accuracy. Further discussion of coherent structures is conducted later in this section.

\begin{figure}
\centering
\includegraphics[width=0.99\textwidth]{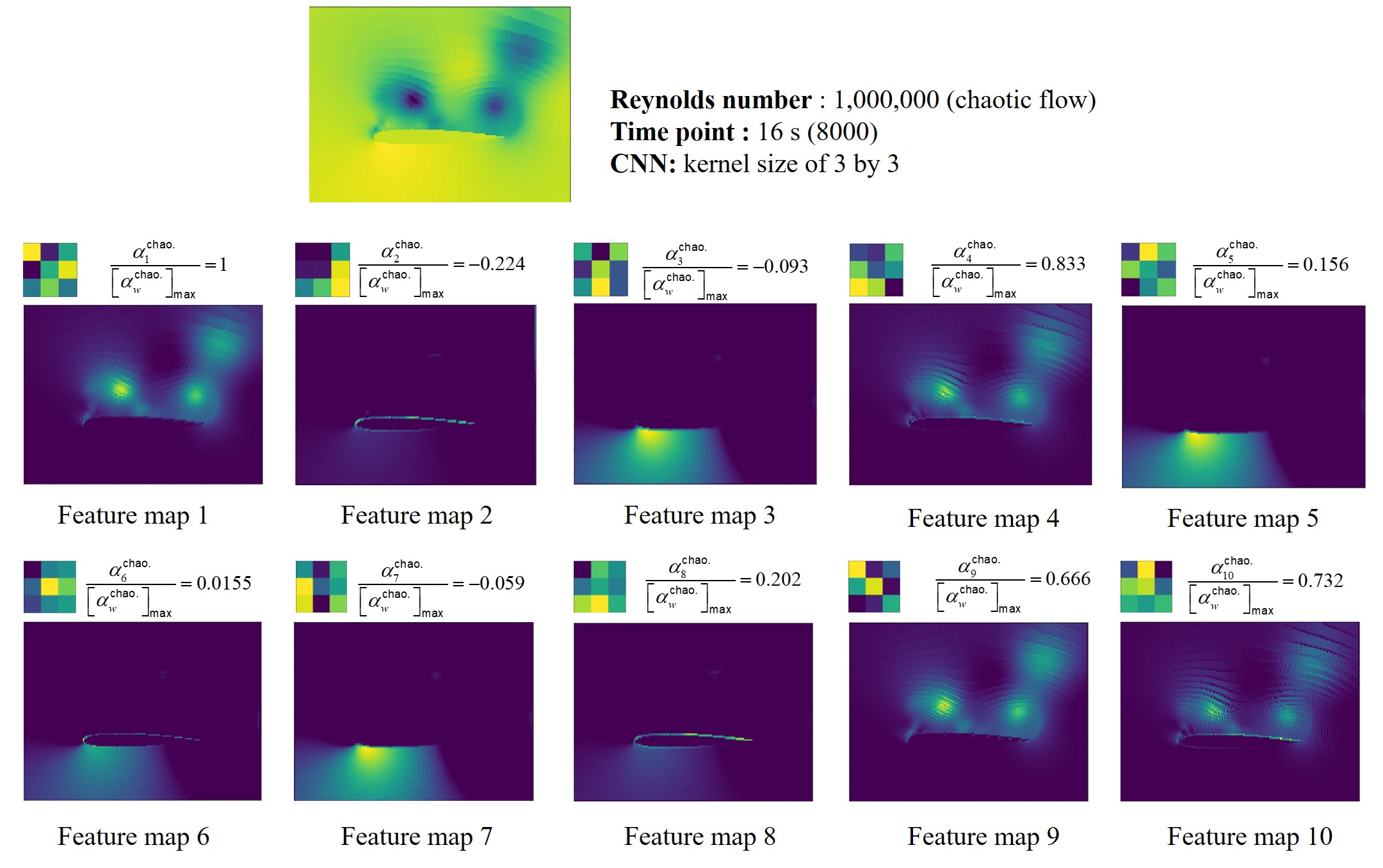}
\caption{\label{fig:5} Ten kernels of the trained CNN and the corresponding feature maps for the example snapshot. The normalized neuron importance weights denotes ${\alpha _{w}^{\mathrm{chao}.}}\Bigg/{\left[ \alpha _{w}^{\mathrm{chao}.} \right]}_{\max}$, where the maximum value $\left[ \alpha _{w}^{\mathrm{chao}.} \right] _{\max}$ is $\alpha _{1}^{\mathrm{chao}.}$.}
\end{figure}

As was discussed in section~\ref{sec: 2.3}, CNNs are known to be particularly successful when the input data can be decomposed into some form of hierarchical basis representation. Those low-dimensional characteristics can be extracted by convolutional kernels and visualized by the feature maps which connect the convolutional layers. Figure~\ref{fig:5} provides an example, where a snapshot of a chaotic flow is taken as the input for the trained CNN (kernel size of 3 pixels square). The feature maps in this plot were representative of all snapshots provided to the neural network. The corresponding convolutional kernel is presented above each feature map. It is clear that each kernel identifies a certain pattern within the pressure field. Feature maps 2 and 8 were the only ones to extract the shape of the airfoil itself; these are thus denoted as “edge kernels.” The 1th, 4th, 9th and 10th kernels accurately extract local, quasi-circular low- and high-pressure regions in the original snapshot, denotes as “bubble kernels.” Feature maps 3, 5 and 7 extracted the high pressure region near the airfoil’s leading edge, and are thus called “high pressure kernels.” 

These results are significant as they show that the neural network automatically identified three large-scale coherent structures without human intervention or knowledge of the flow’s kinematics. This is especially meaningful as only spatial characteristics were provided in each snapshot, without considering the temporal relationship between different snapshots for one flow regime.
The convolutional neural network thus assembled itself as follows. Fundamental patterns were identified at the lowest level which consistently identified the coherent structures characteristic of the qualitatively different flow regimes provided to the model. These coherent structures were then correlated to flow regime classifications in a fully connected high-level layer, to a high degree of fidelity. 

In order to fully understand the importance of these identified patterns for classification, a technique called Gradient-weighted Class Activation Mapping (Grad-CAM)~\cite{selvaraju2017grad} was applied to the trained model for obtaining the neuron importance weights $\alpha _{w}^{c}$
, which captures the importance of feature map $w$ for a target class $c$:
\begin{equation}
\alpha _{w}^{c}={\overset{\text{global\,\,average\,\,pooling}}{\overbrace{\frac{1}{Z}\sum_i{\sum_j{\underset{\text{gradient\,\,via\,\,backprop}}{\underbrace{\frac{\partial y^c}{\partial F_{ij}^{w}}}}}}}}}
\end{equation}

The number of target class $c$ in our model is three (periodic flow $c$: perio., quasi-periodic flow $c$: qua. and chaotic flow $c$: chao.). $\frac{\partial y^c}{\partial F_{ij}^{w}}$ represents the gradient of the score for class $c$, $y^c$, with respect to a pixel $F_{ij}$ in feature maps $w$ of a convolutional layer. $Z$  represents the number of pixels in the feature map. Of note is that only feature map with positive values will be emphasized, larger values in weights means higher importance of these features in classification. Here, we calculated weights of ten feature maps for chaotic flow. The relative weights (also included in Fig.~\ref{fig:5}) of the feature maps indicate that different certain structures were valued more than others for the emergence of chaotic flow. The edge kernels were less valued for the occurrence of buffet flow, which means that the model considered the airfoil shape of little value for determining the type of flow regime. This aligns with thin-airfoil aerodynamic theory, from which it can be concluded that at this high incidence angle the airfoil profile minimally influences the qualitative flow characteristics. The high-pressure kernels were also unvalued for the right classification of buffet flow. All bubble kernels were more heavily weighted than were the other kernels.

Consequently, the presence of localized fluctuations in pressure was found to most significantly inform the model’s flow classification. If the bubbles were present much more than was the high-pressure region near the airfoil’s leading edge, the flow was classified as chaotic. If the bubbles existed with comparable magnitude to the high-pressure region, the flow was quasiperiodic. If the bubbles were much less present than was the high-pressure region, then the flow was periodic. This nuanced classification algorithm, developed entirely by the neural network, aligns with an understanding of the airfoil flow’s kinematics. For example, Kurtulus~\cite{kurtulus2015unsteady} observed through a rigorous analysis of wake structures a similar pattern in coherent structures.

These results were obtained with kernels which were three pixels square; a brief study was conducted to understand the sensitivity of the coherent structure identification to kernel size. The same procedure was followed as outlined above, with only the kernel size changing. Table~\ref{tbl4} details the number of kernels in each dynamical category, as developed above. While the edge kernels do not appear when the kernels are larger than 5 pixels square, more useless kernels, showing little coherence and are thus denoted as “useless kernels”, appear as the kernel size increases. The dynamically valuable kernels, viz. the bubble and the high-pressure kernels, exhibited less sensitivity to kernel size but also became less common as the kernel size increased significantly. Edge detection is known to require smaller kernels, and the lack of dynamical significance of the airfoil edge does not motivate the model to try to retain the airfoil shape information. The pressure bubbles are themselves rarely larger than 20 pixels in diameter with this interpolated resolution, so again it makes sense that in a low convolutional layer the kernels would struggle to take a form which can consistently identify the bubbles. Consequently, the loss of coherent structures with increased kernel size was not surprising.

\begin{figure}
\centering
\includegraphics[width=0.5\textwidth]{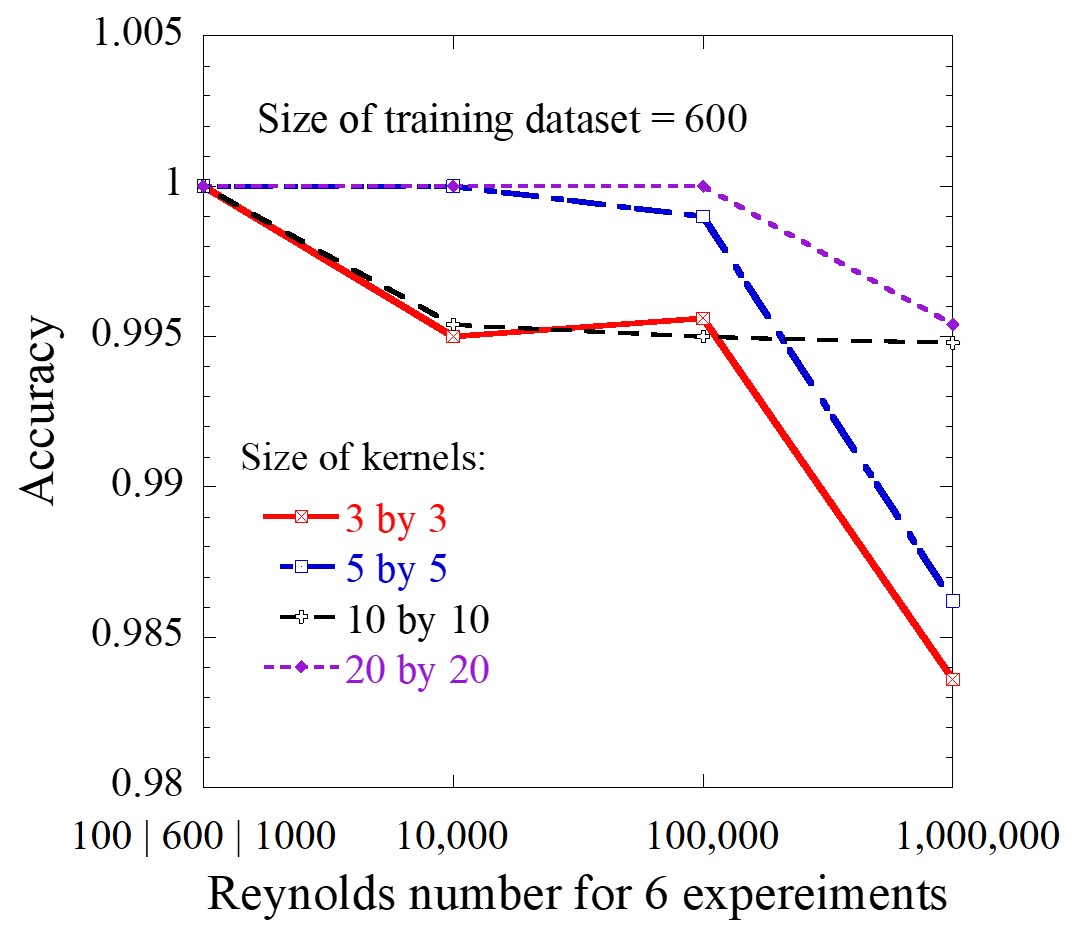}
\caption{\label{fig:6} Test accuracy of 6 different experiments (Reynolds numbers) for 4 CNNs with different kernel sizes.}
\end{figure}

Although the coherent structures were not as clearly identified with the larger kernels, the large-kernel models still performed well in their classification task as summarized in Fig.~\ref{fig:6}. This can be understood by the concept of “receptive field” in ML, which is the region of the input space that affects a particular unit of the network. In our model, we only have one convolutional layer, and the size of kernel is the value of receptive field. When increasing the value of receptive field, the information that neurons can contact is much larger, which means that the kernel can summarize more global information. Corresponding features are much more abstract than those with little kernels, whose information is organized locally and with more detail, and can therefore not be as easily understood by the users. The definition of a useless kernel is simply a kinematic one, corresponding to resulting feature maps which cannot be easily understood by a human as a known dynamically significant pattern. However, these abstract features will be very useful for a computer to distinguish the category, and that is the reason why the accuracy for them is still high.
\begin{table}
\caption{Number of functional kernels for different trained CNNs.}\label{tbl4}
\centering
\begin{tabular}{lllll}
\toprule
Kernel size & Edge Kernel & Bubble kernel & High-pressure kernel & Useless kernel\\
\midrule
3 × 3 & 2 & 4 & 4 & 0 \\
5 × 5 & 1 & 5 & 2 & 2 \\
10 × 10 & 0 & 1 & 2 & 7 \\
20 × 20 & 0 & 0 & 0 & 10 \\
\bottomrule
\end{tabular}
\end{table}

\section{Convolutional Long-Short Term Memory Implementation}
\label{sec: 4}
The conventional CNN identified several static coherent flow structures in several buffet flows such that it would discern with high accuracy the qualitative nature of the flow regime. These coherent structures, while spatially local by virtue of the convolutional layer, are similar to empirical modes identified using other characterization approaches like the proper orthogonal or dynamic mode decompositions (see the supplementary material). A next logical step, therefore, is to introduce temporal information into the neural model so that it can compute characteristic time scales associated with these coherent structures. 

\subsection{Formulation}
\label{sec: 4.1}
The LSTM network was first proposed by Sepp Hochreiter and Jurgen Schmidhuber~\cite{hochreiter1997long} in 1997 as a variant of the recurrent neural network (RNN). A mathematical summary of the LSTM architecture is presented in the supplementary material. It can not only process single data points (such as images), but also entire sequences (such as speech or video). The LSTM architecture improved the capability of processing long data sequences by addressing stability bottlenecks like the vanishing gradient which frustrated early RNN implementations. In section~\ref{sec: 3}, the input data sample for the CNN was a single temporal snapshot; only spatial characteristics were considered in the model. In this LSTM architecture, an untrained CNN was still used to convert the 2D flow snapshots into characteristic 1D vectors, and then these vectors were fed into the LSTM network chronologically. Employing a single CNN in this way differs from conventional LSTM architectures and allows kernels comparable to those in Fig.~\ref{fig:5} to be computed from the CNN-LSTM. Because our quantifiable task is to differentiate between three different flows, only the output of the last cell is desired from the standpoint of training the model; the coherent structures identified along the way remain of fundamental interest.

\begin{figure}
\centering
\includegraphics[width=0.8\textwidth]{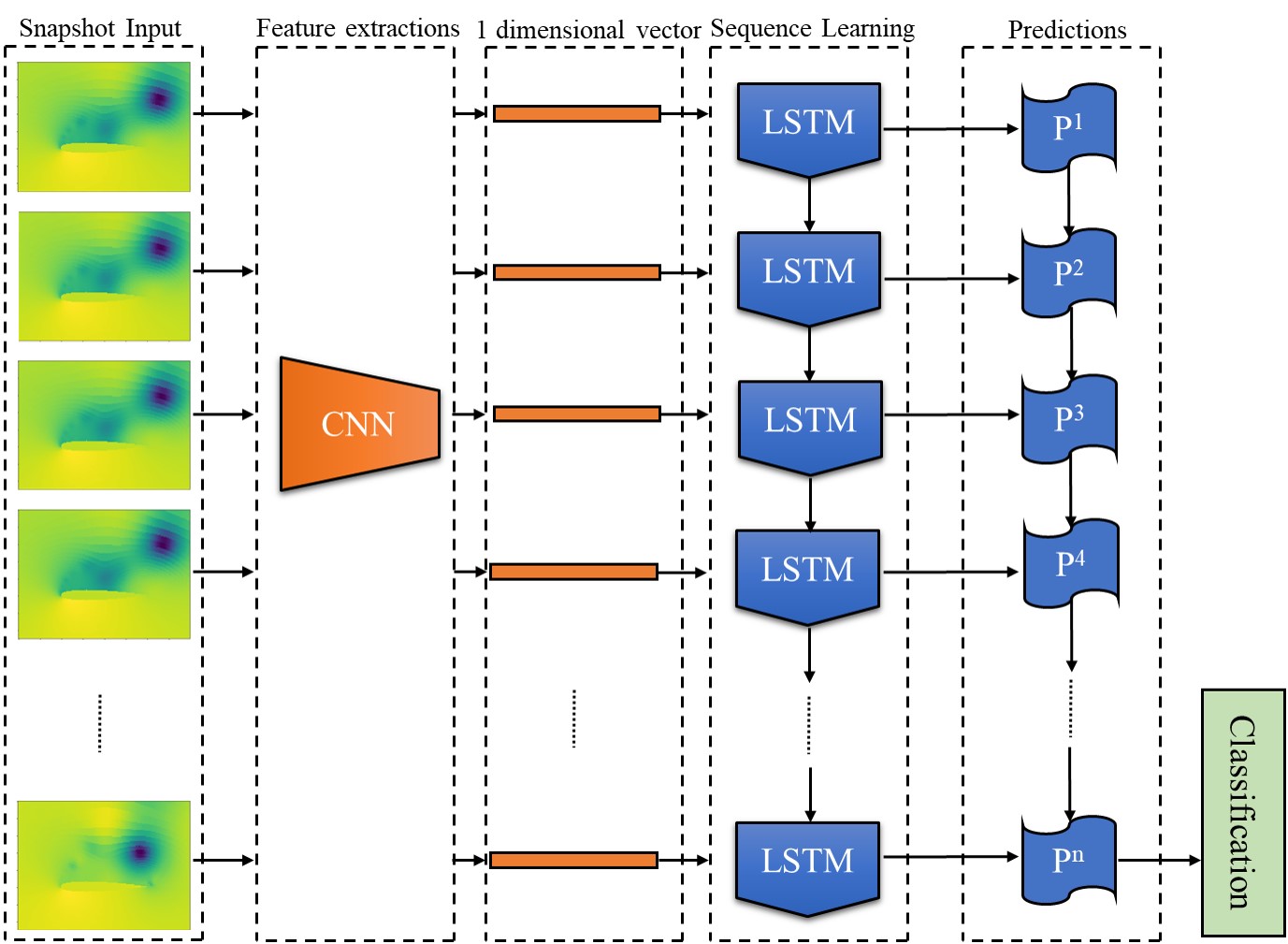}
\caption{\label{fig:7} CNN-LSTM architecture overview.}
\end{figure}

The architecture of the implemented LSTM network is outlined in Fig.~\ref{fig:7}. Hyperparameters for the CNN had the same values as those in Table~\ref{tbl2}, with only kernels of size 3 by 3 being implemented for the results in this section. Other LSTM hyperparameters are outlined in Table~\ref{tbl5}.
\begin{table}
\caption{LSTM hyperparameters.}\label{tbl5}
\centering
\begin{tabular}{ll}
\toprule
Hyperparameter & Value\\
\midrule
Number of units in LSTM & 200 \\
Length of sequence (snapshot) & 20 \\
Batch size & 1 \\
Dropout ratio for LSTM & 0.75 \\
\bottomrule
\end{tabular}
\end{table}

\subsection{Results and Discussion}
\label{sec: 4.2}
\begin{figure}
\centering
\includegraphics[width=0.99\textwidth]{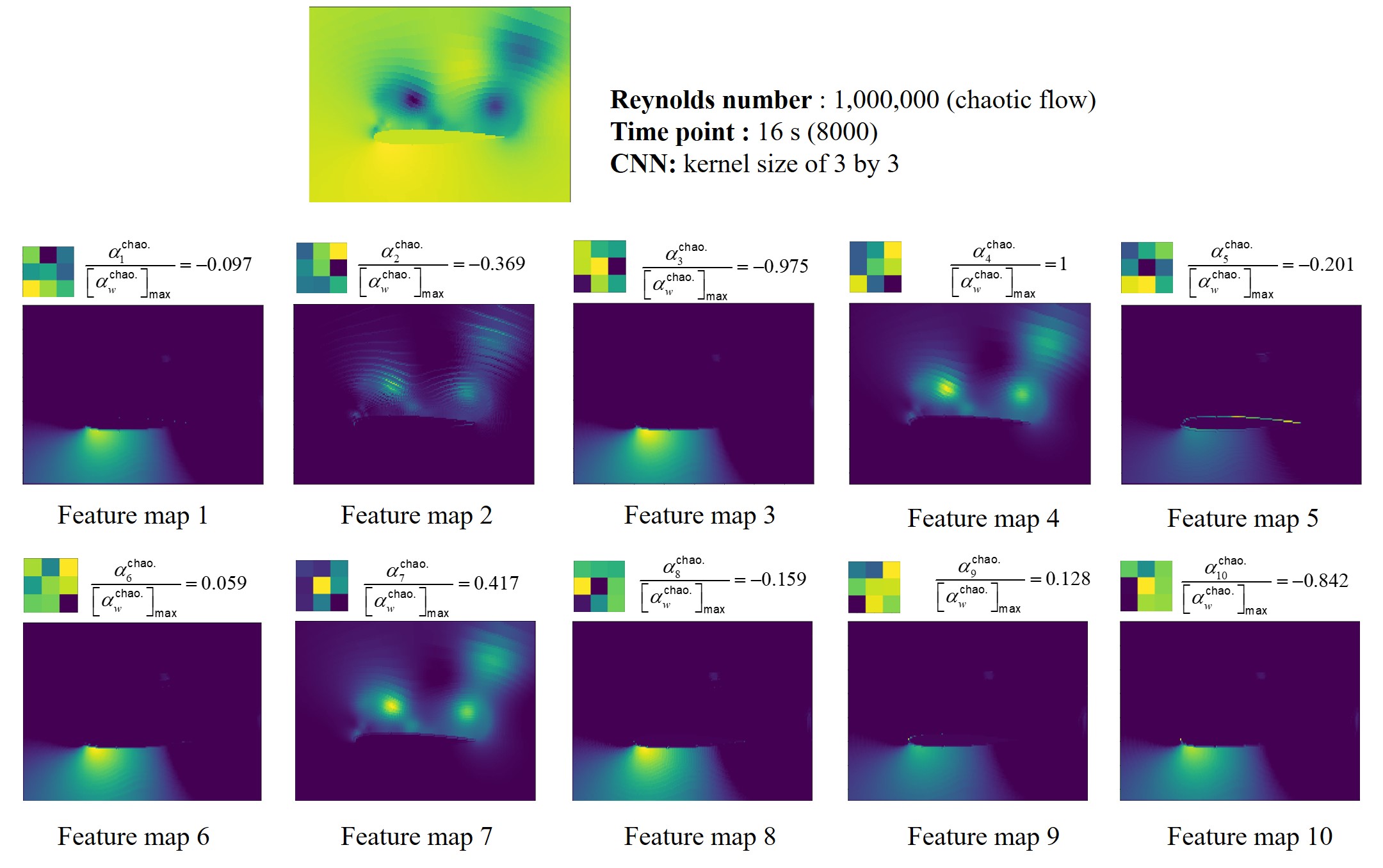}
\caption{\label{fig:8} Ten kernels of the trained CNN-LSTM and the corresponding feature maps for an example snapshot. The number above every feature map is the normalized neuron importance weights.}
\end{figure}

The number of snapshots in one data sequence was set to 20, which means that 600 snapshots in the training dataset were divided into 30 data sequences. These 30 training sequences resulted in a model which was then tested on 1050 test sequences comprising the entire DNS data set outlined in section~\ref{sec: 2}. The resulting classification accuracy was nearly perfect: the machine classified falsely 1 out of 1050 data sequences after 10,000 training steps. Figure~\ref{fig:8} shows kernels and their corresponding feature maps in the convolutional layer for the trained CNN-LSTM. 

Compared with those in Fig.~\ref{fig:5}, it is clear that the CNN also successfully identified three large-scale coherent structures. As was the case with the CNN in section~\ref{sec: 3}, the bubble kernels (kernel 4 and 7) will play a dominant role in the task of classification as their neuron importance weights are much larger than most of other kernels. Of note is that although the kernel 2 also have the tendency to identify the bubbles, its importance in classification is negligent as the result of its less dynamical information.

The CNN-LSTM model thus identified coherent structures similar to those in the conventional CNN when provided with the same training data sets, and performed comparably for the nominal classification task set forth. With the inclusion of temporal information, the CNN-LSTM model have nearly perfect accuracy for test dataset. The trade-off, naturally, is that training a CNN-LSTM involves solving within a much larger parameter space and therefore requires significantly more computational resources to train.

\section{Conclusions and Future Works}
\label{sec: 5}
In this paper, a highly accurate convolutional neural network was successfully trained to recognize different manifestations of subsonic buffet over a high-incidence airfoil when provided with individual temporal snapshots. By extracting convolutional kernels and the corresponding feature maps from the trained model, the capability of identifying large-scale coherent features was validated. Sensitivity of hyperparameters, including the size of the training dataset, convolutional kernel size and general network architecture, were explored. Four main conclusions are stated as follows.
\begin{enumerate}
\itemsep=0pt
\item The trained CNN automatically identified three large-scale structures, including the airfoil edge, localized shedding pressure abnormalities (viz. “bubbles”) and the high-pressure region near the airfoil’s leading edge. This was accomplished without human intervention or knowledge of the flow’s kinematics.
\item The presence of localized fluctuations in pressure (bubbles) was found to inform most significantly the model’s flow classification. These were both highly weighted characteristics in the CNN model and CNN-LSTM model.
\item Smaller convolutional kernels were necessary to identify coherent structures as understandable by human users. Larger convolutional kernels still resulted in highly accurate flow classifications, but were less physically informative to the users due to the “receptive field” concept.
\item Consideration of temporal information in the CNN-LSTM improved the classification accuracy. The multiscale nature of the chaotic flow was identified as dynamically important by the model, again with no provided kinematic information.
\end{enumerate}  

In general, it is demonstrated in this work that the CNN has the potential to extract large-scale coherent structures to achieve specific tasks in complex fluid flows. By applying Grad-CAM technique to the trained model, we can know the relative importance of these identified structures in the task we concerned, which can help us gain further insights into these confounding dynamics from the perspective of machine. This cannot be realized from existing modal decomposition techniques (e.g. POD, DMD). In future work, we believed these “feature-filter” kernels can be preserved for transfer learning to advance our training process in other complex fluid flows. Keeping the parameters in the convolutional layers unchanged while shifting parameters in the full-connected layers can be a promising path. And other flows of engineering interest like jet flows and mixing layers will be studied and other parameters will be varied including the Mach number.

\bibliographystyle{unsrt}  
\bibliography{template}  


\end{document}